\documentclass[twocolumn]{aastex62}
\usepackage{graphics,epsf}
\usepackage{amsmath}                
\usepackage{amsfonts}               
\usepackage{amssymb}                
\usepackage{epsfig}                 
\usepackage{appendix}
\usepackage{graphicx}
\usepackage{float}
\usepackage{color}
\usepackage{multirow}
\usepackage{colortbl}
\usepackage[para,online,flushleft]{threeparttable}

\newcommand{\erg}{{~\rm erg}}
\newcommand{\yr}{{~\rm yr}}

\newcommand{\days}{{~\rm days}}

\begin{document}

\title{Rare events of a peculiar thermonuclear supernova that precedes a core collapse supernova}

\email{ealealbh@gmail.com; soker@physics.technion.ac.il}

\author{Ealeal Bear}
\affiliation{Department of Physics, Technion – Israel Institute of Technology, Haifa 3200003, Israel}

\author{Noam Soker}
\affiliation{Department of Physics, Technion – Israel Institute of Technology, Haifa 3200003, Israel}
\affiliation{Guangdong Technion Israel Institute of Technology, Guangdong Province, Shantou 515069, China}

\begin{abstract}
 We study stellar binary evolution that leads to the formation of a white dwarf (WD) that explodes in a thermonuclear supernova at the termination of a common envelope evolution (CEE) shortly before the core of its companion explodes as a core-collapse supernova (CCSN). The CCSN explosion of the core, which is the remnant of a red supergiant (RSG) star, 
might take place few months to several years after the explosion of the WD as a thermonuclear supernova, i.e., a type Ia peculiar supernova (peculiar SN Ia).  
Using the evolutionary code \textsc{mesa-binary} we simulate evolution of binary systems with stars of initial masses of $6-7.5 M_\odot$. The more massive star, the primary, transfers mass to the secondary star and leaves a CO WD remnant. The secondary becomes massive enough to end in a CCSN. As the secondary evolves to the RSG phase it engulfs the WD and the system experience a CEE that ends with a WD-core binary system at an orbital separation of $a_f \simeq 1-5 R_\odot$. Our simulations show that the core explodes as a CCSN at $t_{\rm CEE-CCSN} \simeq 3000 - 10^5 \yr$ after the CEE.
We assume that if the WD accretes helium-rich gas from the core it might explode as a SN Ia in the frame of the double detonation scenario for SNe Ia and peculiar SNe Ia. We predict the very rare occurrence of a peculiar SN Ia followed within months to years by a CCSN.
\end{abstract}


\section{Introduction} 
\label{sec:intro}

Supernovae (SNe) Ia and some peculiar SNe Ia result from the complete destruction of a white dwarf (WD) in a nuclear burning of carbon-oxygen to nickel, while in some peculiar SNe Ia only part of the WD burns (for reviews from the last couple of years see  \citealt{LivioMazzali2018, Wang2018, Jhaetal2019NatAs, RuizLapuente2019, Soker2019Rev, Ruiter2020}; for very recent new approaches to SNe Ia see, e.g., \citealt{Antoniadisetal2020, Peretsetal2021, Wuetal2021}). 
Because they result from WDs, these thermonuclear SNe are descendant of zero-age main sequence stars with masses of $M_{\rm ZAMS} \la 8.5-9 M_\odot$, as more massive stars end as core collapse SNe (CCSNe; including electron capture SNe). The mass limit between WDs and CCSNe depends on metalicity as well as on some uncertain stellar structure parameters (e.g., \citealt{Hegeretal2003, Poelarendsetal2008, IbelingHeger2013, Dohertyetal2017, GilPonsetal2018}).  
This mass limit implies that in single-evolving stellar populations CCSNe occur much before SNe Ia (including peculiar SNe Ia).

There are binary evolutionary routes that form the WD before they form the NS, e.g, the WD suffers accretion induced collapse and forms a NS (e.g. \citealt{Ruiteretal2019, TaurisJanka2019, LiuWang2020, WangLiu2020}), or a mass-transfer process that brings the initially less massive star to be massive enough to explode as a CCSN (including electron capture; e.g.,  \citealt{TutukovYungelSon1993, PortegiesZwartVerbunt1996, PortegiesZwartYungelson1999, vanKerkwijkKulkarni1999, TaurisSennels2000, Brownetal2001, Nelemansetal2001, Daviesetal2002, Kimetal2003, Kalogeraetal2005, Churchetal2006, vanHaaftenetal2013, Toonenetal2018, Breiviketal2020, BearSoker2021}).  

In this paper we examine some cases where in binary systems a peculiar SN Ia takes place shortly before a CCSN, such that the CCSN explodes inside the peculiar SN Ia remnant. Such an evolutionary route requires the WD to be born before the CCSN that leaves behind a neutron star (NS). This is a {\it WD-NS reverse evolution} (e.g., \citealt{SabachSoker2014}). The general evolution proceeds as follows. The primary, i.e., initially more massive, star starts with a zero age main sequence mass in the range of  $\simeq 5.5 M_\odot - 7.5 M_\odot$. As it evolves and becomes an asymptotic giant branch (AGB) star, it transfers mass to the initially less massive star, i.e., the secondary star. The primary star ends as a CO WD while the secondary star gains mass to $\ga 8.5 M_\odot$. This mass implies that the secondary star might end its life as a CCSN (including an electron capture SN), leaving behind a NS remnant. 
The secondary star is not likely to explode as a CCSN even if it gains mass to $\ga 8.5 M_\odot$ in cases where the mass transfer takes place after the secondary has left the main sequence and developed a helium core (e.g., \citealt{BearSoker2021}).

In this study we follow systems where the secondary star evolves to become a red supergiant (RSG) star and engulfs the CO WD remnant of the primary star and the system enters a common envelope evolution (CEE). \cite{SabachSoker2014} summaries different outcomes of this CEE. The outcome depends among other things on whether the WD is a CO WD or a an ONeMg WD. Here we consider only CO WDs. Some outcomes include the merger of the WD with the core. This merger might lead to a transient event that will be followed by a CCSN, or to a terminal peculiar SN, including peculiar SN II, SN Ibc, or SN Ia (e.g., \citealt{SabachSoker2014, Ablimit2021}). Here we are interested in cases where a thermonuclear supernova takes place shortly before a CCSN.

The thermonuclear explosion of the WD inside the common envelope is a peculiar SN Ia, but most likely be classified as a peculiar CCSN because of the hydrogen-rich common envelope. The CCSN that will follow the thermonuclear explosion be a type IIb since the ejecta will be low in hydrogen ($M_{\rm H} < 0.5 M_\odot$), or even a SN Ib. We specifically are looking for cases where the time from the peculiar SN Ia and the CCSN is short, such that the CCSN takes place inside a SN remnant (SNR). We will not follow the system to the thermonuclear explosion (we do not have the tools for that), but assume that if a thermonuclear explosion occurs it does so shortly after the onset of the CEE or shortly before the CCSN of the core. We are therefore looking for systems where the time from the onset of the CEE to CCSN is $t_{\rm CEE-CCSN} \la 10^5\yr$. Although this scenario is very rare, we nonetheless study some of its properties because very rare SNe might result from such a scenario.

As we propose a new scenario for a CCSN inside a SN~Ia remnant, we first describe it in section \ref{sec:scenario}. This scenario contains phases that we demonstrate in this study (section \ref{subsec:ScenarioSim}), and parts that are more on the speculative side at present and require future studies to confirm them (section \ref{subsec:ScenarioSpec}). In section \ref{sec:simulations} we describe our numerical procedure, and in section \ref{sec:results} we describe our results that demonstrate the phases of our proposed CCSN inside SN~Ia scenario that we describe in section \ref{subsec:ScenarioSim}.  We discuss some observational consequences in section \ref{sec:Observational}.  Our summary and discussion of the required future simulations are in section \ref{sec:summary}.

\section{A CCSN inside a peculiar SN Ia remnant} 
\label{sec:scenario}

\subsection{The overall scenario} 
\label{subsec:ScenarioAll}

\begin{figure*}
\includegraphics[trim=0.0cm 2.0cm 0.0cm 0.0cm ,clip, scale=0.80]{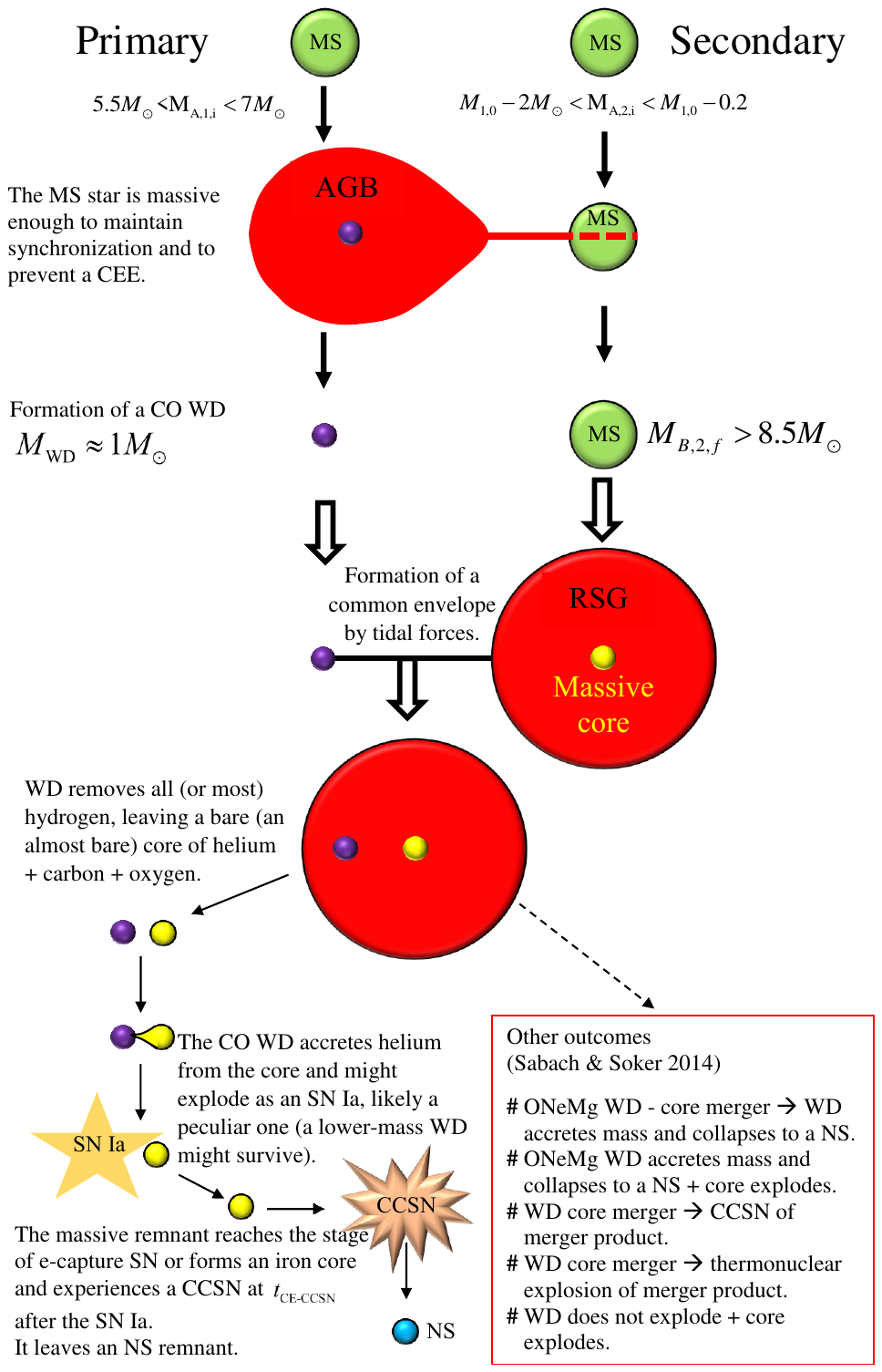}
\caption{A schematic diagram of our proposed evolution towards a rare type Ibc CCSN that takes place inside a remnant of a peculiar SN Ia. The thick arrows depict phases we do simulate: single-line arrows for Stage A with \textsc{mesa-binary} and double-line arrows for Stage B with \textsc{mesa-binary}. We also use \textsc{mesa-single} to calculate the time it would take the secondary to explode as a CCSN (not shown in the figure). Thin solid lines are evolutionary phases of the scenario that we do not simulate in the present study. }
 \label{fig:SchematicScenario}
 \end{figure*}

The very rare scenario of CCSN inside a SN Ia remnant that we study here has some key ingredients and assumptions, as we schematically draw in Fig. \ref{fig:SchematicScenario} and discuss in this section. We simulate the evolutionary phases that we depict by the thick single-line arrows and by the double-line arrows. We describe these phases in section \ref{subsec:ScenarioSim} and describe the results in section \ref{sec:results}. We do not simulate the post-CEE phases that might lead to a peculiar SN~Ia. We only speculate about these phases (section \ref{subsec:ScenarioSpec}).  

\subsection{The phases we simulate} 
\label{subsec:ScenarioSim}

The primary star of initial mass $5.5 M_\odot \la M_{\rm A,1,i} \la 7.5 M_\odot$ evolves to become an AGB star and forms a WD.
The subscripts `A' and `B' refer to the two numerical stages that we describe later, while `1' and `2' refer to the primary star and secondary star, respectively. The subscripts $ {\rm i}$ and ${\rm f}$ stand for the initial and final time of the respective stage. We require the primary star to leave a CO WD remnant, and for that it cannot be more massive than about $7.5 M_\odot$.
We also require that the primary star transfers mass to the secondary star before the secondary leaves the main sequence (otherwise it will not explode as a CCSN), and therefore, the two stars cannot be too close in mass \citep{BearSoker2021}. Crudely, we require 
$M_{\rm A,1,i} - 2 M_\odot \la M_{\rm A,2,i} \la  M_{\rm A,1,i} - 0.2 M_\odot$, where $M_{\rm A,2,i}$ is the initial mass of the secondary star. 
Since the secondary star is massive, it can bring the giant primary envelope to synchronization, and the mass transfer is stable and the system avoids the CEE at this phase. The primary star leaves a CO WD remnant. At the end of the mass transfer process the secondary star has a mass of $M_{\rm B,2,i}$.

The next phase of binary interaction is when the secondary star becomes a RSG and engulfs the CO WD remnant. The system enters a CEE. We consider systems where the CO WD ends at a close orbit $a \approx 1 - 5 R_\odot$ from the core of the RSG star, as we are after CO WDs that explode as (peculiar) SNe Ia. In principle, CO WDs that avoid the CEE and accrete from the wind of the RSG might explode in the frame of the single-degenerate scenario (see reviews cited in section \ref{sec:intro}). We consider the single degenerate scenario with a RSG star that evolves fast to be unlikely (not clear that the WD can gain mass to the Chandrasekhar mass limit $M_{\rm Ch}$), although not zero.

As we demonstrate in section \ref{sec:results}, in the systems that we study here the WD enters a CEE with the RSG of the secondary star, and ends the CEE at an orbital separation of $a_f \simeq 1-5 R_\odot$. We find that the core of the secondary star explodes as a CCSN at a time of $t_{\rm CEE-CCSN} \simeq  3 \times 10^3 - 1.1 \times 10^5 \yr$ after the CEE phase of the binary system. However, we do not simulate the post-CEE binary evolution. Before we present our results of the binary evolution until the CEE phase and then the likely formation of a core-WD close binary system, we complete the description of the scenario by describing the phases of the scenario that require further study. 

\subsection{Possible late evolution toward a peculiar SN Ia} 
\label{subsec:ScenarioSpec}

We start from our finding that at the end of the CEE the WD-core system ends at a separation of about $a_f \simeq 1 -5 R_\odot$, for a CEE efficiency parameter (the fraction of the binary-orbital-energy that goes to unbind the envelope) of $\alpha_{\rm CEE} \simeq 0.2-0.8$. 

Now comes our strongest assumption  for which a detailed rigorous analysis is beyond the scope of this work   because it is more complicated due to the post-CEE core of the RSG star. We assume that at this final orbital separation the WD accretes helium-rich  material from its companion, either from material left over from the CEE or from the core itself by some later Roche lobe overflow.  Based on some other studies (e.g., \citealt{WoosleyWeaver1994, LivneArnett1995, Shenetal2013}) we assume that as the WD accretes helium it might explode as a SN Ia or a peculiar SN Ia in the frame of the double detonation scenario (for a recent study of this scenario see, e.g., \citealt{Mageeetal2021}). More likely, most double detonation explosions lead to peculiar SNe Ia \citep{Soker2019Rev}, e.g., as \cite{Peretsetal2010} suggest for the peculiar SN~2005E, and because the explosion is highly non-spherical (e.g., \citealt{Boosetal2021}), in contradiction with regular SN Ia remnants \citep{Soker2019Rev}. Therefore, in what follows we refer to a peculiar SN Ia. According to the double detonation scenario of SNe Ia, a CO WD accretes helium to the degree that the helium layer detonates and in turn explodes the entire CO WD. Later, in the scenario we study here, the core (remnant of the RSG) explodes as a CCSN. As there is no hydrogen-rich envelope at this post-CEE phase, the CCSN is of type Ib or even of type Ic if at this stage there is no helium left in the system.  

We emphasize that this scenario is very delicate, and so very rare, with the following critical challenges. ($i$) The post-CEE orbital separation should be in the right value. This is the least severe challenge as we show in this paper. The following two challenges must be overcome by future studies. ($ii$) The WD should accrete helium-rich gas, about $0.01 - 0.1 M_\odot$, such that it explodes. A general problem to the double detonation scenario is that it is not clear that the detonating helium explodes the CO WD (e.g., \citealt{Pakmoretal2021}). ($iii$) We require that the WD explosion does not destroy the core of the RSG (as we need the core to experience a CCSN explosion). In some cases the helium-donor does not survive, either after the CO WD explodes and sets an explosion in the helium-donor (e.g., \citealt{Papishetal2015}; a triple detonation scenario), or after the helium detonates and explodes the mass-donor star rather than exploding the mass-accreting CO WD \citep{Pakmoretal2021}.

The helium transfer from the core to the WD, and therefore the thermonuclear explosion of the WD as a peculiar SN Ia, might take place at one of two different time periods.

\begin{enumerate}
\item \textit{Early thermonuclear explosion.} In this case  helium transfer takes place at the end of the CEE or shortly after that as the spiraling-in process brings the WD sufficiently close to the core to accrete mass. Namely, the time delay from the thermonuclear explosion to the CCSN explosion, $t_{\rm Ia-CCSN}$, is about equal to the time delay from the CEE to the CCSN explosion $t_{\rm Ia-CCSN} \simeq t_{\rm CEE-CCSN}$.
This implies that the thermonuclear explosion takes place while there is still a hydrogen-rich envelope or, if the binary ejected all the hydrogen-rich envelope, a close circumbinary matter. The explosion will be classified as a peculiar SN II, e.g., peculiar by the large amount of nickel, although it is a thermonuclear explosion. In some cases the later CCSN will interact with an old SNR, and the region around the CCSN might be quite empty from gas as the SN Ia ejecta cleaned that region.  
\item \textit{Late thermonuclear explosion.} In this case the post-CEE orbital separation is too large for a mass transfer. The systems continues to evolve as two separate stars, a CO WD and a CCSN progenitor (the core). At a time of $t_{\rm CEE-CCSN}$ minus few years, namely, few years before the CCSN, the core (the CCSN progenitor) might expand to the degree that it overflows its Roche lobe and transfers helium-rich gas to the WD. If this process take place, then $t_{\rm Ia-CCSN} \approx {\rm several} \yr$. We might observe two supernovae from the same place with a separation of several years (or even only few months). There are indications that many CCSN progenitors experience enhanced mass loss rate and/or increase in luminosity tense of years to few months before the CCSN explosion itself (e.g., \citealt{Strotjohannetal2021} for a recent paper). These pre-CCSN activities might be due to core activity that excite waves (e.g., \citealt{QuataertShiode2012, ShiodeQuataert2014, Fuller2017, FullerRo2018, Morozovaetal2020}) or by core-magnetic activity (e.g., \citealt{SokerGilkis2017}).
These activities, waves or magnetic activity, might cause the envelope to expand (e.g., \citealt{McleySoker2014}). This expansion can lead the core to transfer helium-rich mass to the WD, leading to a peculiar SN Ia just years before the CCSN. \end{enumerate}

As said, the process of helium transfer to the WD will be a subject of a future study.

\section{Numerical Setup} 
\label{sec:simulations}
\subsection{General overview} 
\label{subsec:General}

We use \textsc{mesa-binary} (Modules for Experiments in Stellar Astrophysics; version 10398; \citealt{Paxtonetal2011, Paxtonetal2013, Paxtonetal2015, Paxtonetal2018, Paxtonetal2019}) to simulate some of the phases in the scenario that leads to a CCSN explosion inside a SN Ia remnant. 
Conducting these simulations with \textsc{mesa-binary} is not straightforward. \cite{GibsonStencel2018} who modeled the evolutionary state of the interacting binary epsilon Aurigae, indicated the difficulties of such simulations (runs).  
Therefore, we divide the evolution into three stages. 

In the first stage, Stage A, we use \textsc{mesa-binary} and follow both stars until the He mass of the primary star decreases to a value below $M_{\rm A,1,f}{\rm [He]}=0.05M_\odot$, mainly due to mass transfer to the secondary star. At the end of Stage A the mass of the secondary star is $M_{\rm A,2,f}=M_{\rm B,2,i} \ga 8.5 M_\odot$, which is its initial mass when we start the second stage, Stage B. 
In Stage B we treat the primary star, which is now a WD, as a point mass (i.e., we do not follow its evolution), but we do use \textsc{mesa-binary} to follow the evolution of the secondary star and of the circular obit. We take the mass of the WD in Stage B to be only the CO core mass, as a fast wind will remove most of the remaining hydrogen and helium rich envelope. We end stage B when the system enters a CEE after the secondary expands as a RSG star (Fig. \ref{fig:SchematicScenario}). We cannot follow the CEE with the tools at hand. For that, to determine the time from the onset of the CEE until CCSN of the secondary star we use \textsc{mesa-single}. We terminate Stage C when oxygen burns in the core (as the time to explosion is very short). We emphasize that Stage C is a numerical bypass of the CEE, and it does not take place in the scenario we study here.

In Stage A, the initial parameters are $M_{\rm A,1,i}$, $M_{\rm A,2,i}$ and $a_{\rm A,i}$, i.e., the primary mass, the secondary mass, and the initial semi-major axis (orbital separation in our circular orbit), respectively. For numerical reasons, mainly too short time steps that make the simulation too long (see \citealt{BearSoker2021}) we stop this stage when the He mass of the primary is below $M_{\rm A,1,f}{\rm [He]}=0.05 M_\odot$, and it clearly will form a WD.  

The simulation is relevant to our study if during Stage A the the mass transfer brings the secondary star to a mass of  $M_{\rm B,2,i}=M_{\rm A,2,f} \ga 8.5 M_\odot$. The initial binary orbital separation (in this study we consider only circular orbits) of Stage B is the orbital separation at termination of Stage A, $a_{\rm B,i} = a_{\rm A,f}$.  
 
The initial primary mass at Stage B is the core mass at the end of Stage A, i.e. $M_{\rm B,1,i}$=$M_{\rm A,1,f}-M_{\rm A,1,f} {\rm [H]}-M_{\rm A,1,f} {\rm [He]}$, where $M_{\rm A,1,f} {\rm [H]}$ and $M_{\rm A,1,f} {\rm [He]}$ are the mass of hydrogen and Helium of the primary star at the end of Stage A respectively. We stop the \textsc{mesa-binary} simulation of Stage B when either the radius of the secondary star exceeds the binary orbit, $a_{\rm B,f}<R_{\rm B,2,f}$, as at this time the system enters a CEE, or when oxygen burns, as it implies that a CCSN explosion takes place instead of a CEE. 

As we cannot follow the CEE, in Stage C we only determine the secondary star evolution time to CCSN by using \textsc{mesa-single}. The initial secondary star is that at the end of Stage A, i.e., is $M_{\rm C,2,i}=M_{\rm A,2,f}$. We evolve the secondary star from the termination of Stage A up to the formation of the CCSN (or very close to it). We take the time of just approaching a CCSN by either one of following two conditions (similar to what we did in \citealt{BearSoker2021}).
(1) $\log(T_c/K) >9.1$, where $T_c$ is the core temperature; or  (2) $\log(L_{\rm nuc}/L_\odot) > 10 $ where $\log(L_{\rm nuc})$ is the total power from all nuclear reactions. These conditions imply that oxygen burns in the core. 

\subsection{Numerical details} 
\label{subsec:details}

All runs in the \textsc{mesa-binary} mode are mostly based on the binary parameters from the \textit{inlists} of \cite{GibsonStencel2018}. We use the mass-transfer scheme of \cite{KolbRitter1990} and mass-transfer efficiency scheme as in \cite{GibsonStencel2018} (from \citealt{Sobermanetal1997}), similar to what we did in \cite{BearSoker2021}. We take the fractional mass-loss from the vicinity of the donor star, lost in a fast wind, as $\alpha=0.1$, and the fractional mass-loss from the vicinity of the accretor star, also lost in a fast wind, as $\beta=0.1$. For the fractional mass-loss from the circumbinary coplanar toroid we take $\delta = 0.1$, where the radius of the circumbinary coplanar toroid is equal to $\gamma^2 a$, where $a$ is the binary semi-major axis. We further follow \cite{GibsonStencel2018} and adopt $\gamma = 1.3$. 
For both stars our initial equatorial surface rotation velocity is $v_{\rm A,1,i} = v_{\rm A,2,i} =2~{\rm km}~{\rm s}^{-1}$ as in \cite{GibsonStencel2018} and the metallicity is set initially (Stage A) to 0.01 as in \cite{GibsonStencel2018}. 

The initial conditions and termination limits are explained in subsection \ref{subsec:General}. Other \textit{inlist} parameters that do not follow \cite{GibsonStencel2018} are as follows. 
\begin{itemize}
\item We consider only circular orbits, and in the simulations we present in section \ref{sec:results} we take into account tidal forces ($do~tidal~sync = true$) for Stages A and B.
   
\item For each star we include a wind according to the Dutch prescription given in \textsc{meas~single} (for more details see \citealt{MaederMeynet2001,Glebbeeketal2009,Vinketal2001,NugisLamers2000}). The wind scaling factor is $\eta=0.8$.

\item For the nuclear reaction network, we take the \textsc{mesa} default. 

\item In order to avoid convergence issues in Stages B and C (especially in the He flash stage), we do not limit the minimum time step in all runs (we use $10^{-100} {\rm s}$ as the lowest boundary, similar to \citealt{GofmanSoker2020}). 

\item {\it varcontrol~target} is set to the default of \textsc{mesa} (this is the target value for relative variation in the structure from one model to the next).
\end{itemize}

Other parameters that we do not mention here are set to their default option in \textsc{mesa-binary} version 10398.

\section{Results}
\label{sec:results}
\subsection{Relevant simulations}
\label{subsec:Simulations}
  
We present our \textsc{mesa} simulations of the binary evolution in Table \ref{tab:outcome}.

\begin{table*}[]
\centering
\begin{tabular}{|l|l|l|l|l|l|l|l|l|l|l|l|l|l|}
\hline
\multirow{2}{2em}{Line No.} & & & &   \multicolumn{10}{c|}{Simulation numerical identifier}    \\ 
 & Stage & {Parameter} & Units & R1 & R2 & R3 & R4 & R5 & R6 & R7 & R8 & R9 & R10 \\ \hline
1 & A & $M_{\rm A,1,i}$ & $M_\odot$ & 7.5 & 7.5 & 7.5 & 7.5 & 7.5 & 7.5 & 7.0 & 7.0 & 7.0 & 7.0 \\ \hline
2 & A & $M_{\rm A,2,i}$ & $M_\odot$ & 7.0 & 7.1 & 7.1 & 7.1 & 7.2 & 6.5 & 6.5 & 6.0 & 6.3 & 6.3 \\ \hline
3 & A & $a_{\rm A,i}$ & $\days$ & 50 & 70 & 60 & 50 & 50 & 50 & 70 & 50 & 70 & 50 \\ \hline
4 & A & $M_{\rm A,1,f}$ & $M_\odot$ & 1.00 & 1.06 & 1.03 & 1.00 & 1.00 & 0.98 & 1.02 & 0.96 & 1.01 & 0.96 \\ \hline
5 & A & $M_{\rm tot} [\rm C]$ & $M_\odot$ & 0.45 & 0.47 & 0.46 & 0.45 & 0.45 & 0.44 & 0.46 & 0.43 & 0.46 & 0.43 \\ \hline 
6 & A & $M_{\rm tot} [\rm O]$ & $M_\odot$ & 0.49 & 0.52 & 0.50 & 0.49 & 0.49 & 0.48 & 0.50 & 0.46 & 0.50 & 0.47 \\ \hline \hline
7 & B & $M_{\rm B,2,i}$ & $M_\odot$ & 11.37 & 11.51 & 11.47 & 11.46 & 11.56 & 10.89 & 10.60 & 10.10 & 10.42 & 10.39 \\ \hline
8 & B & $M_{\rm B,2,f}$ & $M_\odot$ & 9.50 & 9.88 & - & 9.74 & 9.57 & 9.36 & 8.99 & 9.10 & - & 8.94 \\ \hline
9 & B & $R_{\rm B,2,i}$ & $R_\odot$ & 4.39 & 4.37 & 5.30 & 4.32 & 4.38 & 3.77 & 4.11 & 3.81 & 3.93 & 4.01 \\ \hline
10 & B & $R_{\rm B,2,f}$ & $R_\odot$ & 519 & 616 & - & 524 & 530 & 478 & 556 & 427 & - & 534 \\ \hline
11 & B & $a_{\rm B,i}$ & $R_\odot$ & 821 & 978 & 901 & 831 & 841 & 768 & 892 & 701 & 868 & 868 \\ \hline
12 & B & $M_{\rm tot} [\rm He]$ & $M_\odot$ & 4.00 & 3.98 & - & 4.13 & 4.12 & 3.96 & 3.62 & 3.55 & - & 3.59 \\ \hline
13 & B & $M_{\rm tot} [\rm C]$ & $M_\odot$ & 0.61 & 0.57 & - & 0.53 & 0.72 & 0.44 & 0.48 & 0.49 & - & 0.47 \\ \hline
14 & B & $M_{\rm tot} [\rm O]$ & $M_\odot$ & 1.19 & 1.21 & - & 1.17 & 1.14 & 0.97 & 1.08 & 0.82 & - & 0.98 \\ \hline
15 & B & $M_{\rm core} [\rm He]$ & $M_\odot$ & 3.79 & 3.71 & - & 3.67 & 3.93 & 3.25 & 3.36 & 2.77 & - & 3.20 \\ \hline \hline
16 & B & $t_{\rm WD-CEE}$ & $10^6 \yr$ & 5.41 & 5.47 & - & 5.89 & 5.56 & 7.93 & 7.068 & 8.26 & - & 7.37 \\ \hline
17 & C & $t_{\rm WD-CCSN}$ & $10^6 \yr$ & 5.50 & 5.48 & 2.77 & 5.96 & 5.67 & 8.02 & 7.073 & 8.28 & 8.02 & 7.38 \\ \hline
18 & B-C & $t_{\rm CEE-CCSN}$ & $10^4 \yr$ & 8.20 & 0.290 & - & 7.41 & 11.26 & 9.32 & 0.467 & 1.91 & - & 0.570 \\ \hline
19 & CEE &  &  & Yes & Yes & No & Yes & Yes & Yes & Yes & Yes & No & Yes \\ \hline
20 & B & $E_{\rm bind}$ & $10^{48}\erg$ & 1.83 & 1.32 &  & 2.17 & 2.62 & 1.70 & 1.21 & 1.27 &  & 1.25 \\ \hline
21 & B & $a_f(\alpha_{\rm CEE=0.2})$ & $R_\odot$ & 0.79 & 1.16 &  & 0.69 & 0.57 & 0.83 & 1.11 & 0.97 &  & 1.00 \\ \hline
\end{tabular}
\caption{The relevant simulations of this study  that we mark by the simulation numerical identifier R1-R10.  The subscripts 1 and 2 stand for the primary star (initially more massive star; see Fig. \ref{fig:SchematicScenario}) and for the secondary star, respectively. The subscripts $ {\rm i}$ and ${\rm f}$ stand for the initial and final time of the respective stage.  In all cases the secondary star ends in a CCSN.  
Our binary input parameters are the initial primary mass $M_{\rm A,1,i}$, the initial secondary mass $M_{\rm A,2,i}$, and the initial orbital separation $a_{\rm A,i}$. The time interval $t_{\rm WD-CEE}$ (row 16) presents the time from when the primary star forms a WD to the time when the system enters a CEE, i.e., when the radius of the secondary star in Stage B (row 10) becomes larger than the orbital separation, $R_{\rm B,2,f}> a_{\rm B,f}$. The time interval $t_{\rm CEE-CCSN}$ (row 18) is the time from the onset of the CEE to the CCSN explosion of the secondary star.}
\label{tab:outcome}
\end{table*}

In Stage A (rows 1-6) we follow the evolution until the helium mass of the WD remnant of the primary star is below $M_{\rm A,1,f} {\rm [He]}=0.05 M_\odot$. This includes the phase of high mass transfer rate that occurs at $t_{\rm MT} \simeq 3-4 \times 10^7 \yr$ (measured from the zero age main sequence). We find that at this time the secondary star did not develop a helium core yet. 
In Stage B we simulate the evolution of the binary system concentrating on the secondary star and the orbit, neglecting the evolution of the WD (we treat it as a point mass). We terminate Stage B when the system enters a CEE. We explain the subscripts of the different variable in the table caption.

We evolve Stage B (rows 7-15) by taking the remnant of the primary star to be a point mass of $M_{\rm B,1,f} = M_{\rm A,1,f}{\rm [CO,core]}$. Namely, the WD progenitor loses its H-rich and He-rich envelope layers and the final WD mass is its CO core.

Rows 1-3 in Table \ref{tab:outcome} list our input parameters of the binary system, the two initial masses and the radius of the circular orbit, respectively. We then evolve the binary system through mass transfer until the formation of the WD progenitor by the primary star. 
In row 4 we list the final mass of the WD when the total Helium mass decreases to $0.05M_\odot$, while rows 5 and 6 present the total mass of carbon and oxygen in the WD remnant at that time. The total mass of neon and magnesium at this point is negligible and we do not give it. 

Row 7 presents the initial mass of the secondary star at Stage B, which by our scheme equals the final secondary mass of Stage A $M_{\rm B,2,i}=M_{\rm A,2,f}$. Row 8 presents the final mass of the secondary star in Stage B, i.e., when the CEE starts (the secondary star lost some mass by a wind). Rows 9 and 10 present the initial and final radii of the secondary star in Stage B, respectively. 
Row 11 presents the initial orbital separation of Stage B. The final orbital separation is shortly after the WD entered the secondary RSG envelope. 
Rows 12 -14 represent the total mass of He, carbon and oxygen when the secondary ended Stage B, i.e., entered the CEE.
Row 15 presents the He core mass at the onset of CEE. 
Row 16 gives the time interval from the formation of the WD remnant of the primary star (when its He mass decreases below $0.05 M_\odot$) to the onset of the CEE, $t_{\rm WD-CEE}$. 

In row 18 we calculate the time interval from the formation of a CEE to the CCSN of the secondary star, $t_{\rm CEE-CCSN}$. Because our simulation of Stage B ends at the onset of the CEE, we calculate the value of $t_{\rm CEE-CCSN}$ as follows. 
We conduct a simulation of the evolution of the secondary star with \textsc{mesa-single} to find the time from the WD formation until the explosion of the secondary star $t_{\rm WD-CCSN}$ (row 17). We term this Numerical-Stage C, although this does not take place in reality, it is only a numerical stage to determine the value of $t_{\rm CEE-CCSN}$. We then calculate $t_{\rm CEE-CCSN}$=$t_{\rm WD-CCSN}-t_{\rm WD-CEE}$ (row 18). 

Row 19 depicts if the system enters a CEE or not.
If the system enters a CEE we calculate the expected final orbital separation of the core-WD system. 
We calculate the binding energy of the RSG secondary envelope from $r=1R_\odot$ (about equal to the final orbital separation) to the surface by integrating over the gravitational energy of the envelope and taking half that value by the virial theorem. We note that there is the uncertainty of the binding energy resulting from the not well determined fraction of the recombination energy that goes into envelope ejection. We present the envelope binding energy in row 20. 
We then calculate the post-CEE (final) orbital separation $a_f$ by taking the CEE efficiency parameter to be $\alpha_{\rm CEE}=0.2$, as we do not expect a high efficiency for a WD companion. We do not get into the very long discussion of the $\alpha$-CEE prescription and its large uncertainties. One such large uncertainty is whether there is a post-CEE circumbinary disk that might lead to further shrinkage of the orbit (e.g., \citealt{KashiSoker2011}). Our point is only that the WD might end at a post-CEE orbital separation of $a_f \simeq 1-5 R_\odot$. We list the values of $a_f$ in row 21. 

\subsection{Evolution on the HR diagram}
\label{subsec:Evolution}

\begin{figure*}
\includegraphics[trim=0.0cm 2.0cm 0.0cm 0.0cm ,clip, scale=0.80]{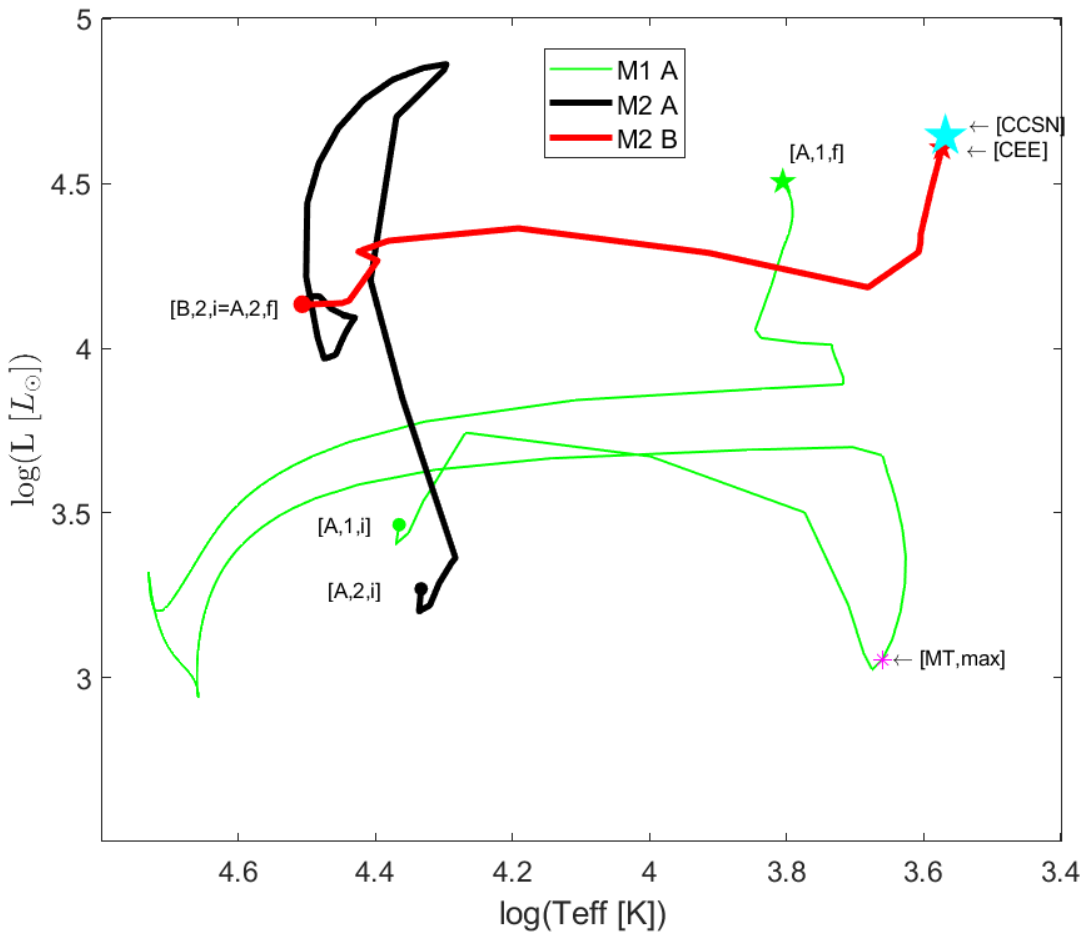}
 \vskip -3.00 cm
\caption{The HR diagram for the case with initial masses and orbital separation of $M_{\rm A,1,i}=7.5 M_\odot$, $M_{\rm A,2,i}=6.5 M_\odot$, and $a_{\rm A,i}=50\days$, respectively. We mark points with the nomenclature that we use throughout the paper. The thin-green line represents the primary star up to the upper AGB (this is the numerical Stage A). We do not follow the evolution to the WD cooling track in this graph. In Stage B the primary is a WD, and we take it to be a point mass in simulating Stage B. The thick-black line represents the secondary star evolution during Stage A. We terminate Stage A when the primary is about to move to the left as its envelope becomes very light ($M_{\rm A,1,f} [He] <0.05$). The thick-red line represents the secondary star evolution during Stage B that we terminate when the binary system enters a CEE. We present the initial points of Stage A, one of the primary and one of the secondary, by green and black dots, respectively. The initial point of the secondary in Stage B (which is the termination point of Stage A) is presented with a red dot. The formation of a CEE is presented by a red pentagram. The formation of a CCSN as defined by the oxygen burn (for more details see Sec. \ref{subsec:General}) is presented by a cyan pentagram. However, we expect the CEE (that we do not include in our simulations) to expose the core and move the secondary star to the left in the HR diagram before it explodes. The maximum mass transfer is presented by a magenta star $[{\rm MT, max}]$.}.
 \label{fig:HR_Diagram_7.5_6.5_50}
 \end{figure*}

In Fig. \ref {fig:HR_Diagram_7.5_6.5_50} we present the evolution of the two stars of simulation R6 on the HR diagram. As our main goal in this study is to explore the evolution towards a possible SN Ia that precedes a CCSN, we do not analyse the early evolution, e.g., the mass transfer. For more details on the mass transfer and the evolution on the HR diagram of WD-NS reverse evolution we point the reader to our earlier paper \cite{BearSoker2021}.
As in Stage B we do not evolve the primary star, we show only the evolution of the secondary star in Stage B. 

Most relevant to the preset study are the following evolutionary properties. 
(1) The CCSN occurs shortly after the CEE, and so our numerical simulations show only small movement of the secondary star on the HR diagram between the CEE and CCSN. However, we recall that we do not simulate the CEE. We expect that due to energy deposition during the CEE the secondary will expand, will become brighter, and will stay very cool. Later, because of mass loss, its core might become exposed and the effective temperature will increase and so the secondary star will move to the left on the HR diagram. This further evolution is beyond our capabilities. 
(2) The high mass transfer rate from the primary to the secondary star occurs relatively rapidly at $t_{\rm MT, max}= 3.16 \times 10^7\yr$. At that time the secondary star is still on the main sequence, and did not develop a Helium core, implying that it will explode as a CCSN (see \citealt{BearSoker2021}). Due to the mass transfer the secondary star becomes very bright ($\approx 10^5 L_\odot$), but later it settles to the position on the main sequence that corresponds to its higher mass, and continues evolution from there.  
 
\subsection{Thermonuclear to CCSN explosion delay time}
\label{subsec:TimeDelay}

There are three main findings from the ten stellar evolutionary simulations that we perform. (1) In eight cases the binary systems experience a CEE and the WD-core binary system ends at a small orbital separation of $a_f \simeq 0.6 - 1.1 (\alpha_{\rm CEE}/0.2) R_\odot$. (2) The WD remnant is of mass $\simeq 1 M_\odot$. (3) The delay time from the CEE to the CCSN is in the range of $t_{\rm CEE-CCSN} \simeq 3000 - 1.1 \times 10^5 \yr$.   

Interestingly, this range of delay times from the CEE to the CCSN explosion includes similar values to the $\approx 20,000 \yr$ from the formation of the three circumstellar rings of SN~1987A to its explosion (e.g.,  
\citealt{CrottsHeathcote2000}). These rings most likely formed as the progenitor of SN~1987A entered a CEE with a main sequence companion that did not survive the CEE. Indeed, some properties of SN~1987A suggest that its progenitor engulfed a main sequence companion (e.g., \citealt{ChevalierSoker1989, Podsiadlowskietal1990}).

According to our proposed scenario in some cases the WD remnant of the primary star accretes helium-rich gas from the core of RSG and experiences a thermonuclear explosion like SNe Ia or peculiar SNe Ia. 
At a time delay of $t_{\rm Ia-CCSN}$ the core of the RSG explodes as a CCSN.
As we discussed in section \ref{subsec:ScenarioSpec} there are two possibilities (and some uncertainties), an early thermonuclear explosion and a late thermonuclear explosion. 
 
The early thermonuclear explosion occurs at the end of the CEE or shortly after it. Due to the hydrogen-rich circumbinary gas and possibly some leftover envelope, the explosion will be classified as a peculiar CCSN. In that case the delay from thermonuclear explosion to CCSN is $t_{\rm Ia-CCSN} \simeq t_{\rm CEE-CCSN} \simeq 3000 - 1.1 \times 10^5 \yr$. 
The CCSN ejecta might expand into an empty interstellar medium bubble that the thermonuclear explosion has cleared thousands to tens of thousands of years earlier. In a case of a peculiar-like SN Ia that leaves some of the WD intact, the CCSN might take place with a WD companion near it. The CCSN will be classified as type Ib or type Ic.  
  
The late thermonuclear explosion  is a speculative proposal.  It  might  takes place only after the core expands to transfer mass to the WD, a process that,  if occurs,  is likely to take place several months to several years before CCSN explosion, i.e., $t_{\rm Ia-CCSN} \approx {\rm several} \yr$ (section \ref{subsec:ScenarioSpec} for more details). This very rare  and speculative  event is of a peculiar SN Ia and a CCSN that occurs few years later. Due to the helium in the system the CCSN will be classified as a type Ib CCSN. 

In a future study we will study the event rate of CCSN inside a peculiar SN Ia, as well as of other WD-NS reverse evolution scenarios.

\section{Some observational consequences}
\label{sec:Observational}
  
 New transient surveys,  e.g., The Zwicky Transient Facility (ZTF; \citealt{Bellmetal2019}), the Large Synoptic Survey Telescope (LSST; \citealt{Ivezicetal2019}), the Southern Hemisphere Variability Survey (LSQ; \citealt{Baltayetal2013}), and the All-Sky Automated Survey for Supernovae (ASAS-SN; \citealt{Kochaneketal2017PASP}), will detect over $10^4$ CCSNe and similar bright transients per year. These very large numbers imply that these surveys will detect very rare events. The many very rare transient events can add up to a significant number of events that do not belong to the common SN classes. For that, the exploration and study of the very rare transient events will allow better classification of the hundreds very puzzling events out of $\ga 10^5$ events that the surveys will find over the coming years. Our proposed scenario of a peculiar SN Ia that precedes a CCSN belongs to the class of very rare transient events, and hence the importance of introducing it.  

 These transient surveys allow observations of early light curves (e.g., \citealt{Maundetal2021}).  \cite{Clarketal2020}, for example, present the early light curve and spectra of SN LSQ13ddu that displayed a rapid rise to maximum. This is a stripped (hydrogen poor) CCSN whose ejecta interacted early on with a circumstellar matter (CSM). In the scenario we study here, the final explosion of the core will be of a stripped CCSN because the thermonuclear explosion of the WD will remove most or even all of its left-over hydrogen-rich envelope, e.g., like in the simulations of \cite{Hiraietal2020} for a somewhat different setting. However, because the thermonuclear explosion of the WD cleaned a large volume there will be no signatures of ejecta-CSM interaction when the core explodes.   

 Some observations categorise supernova remnant types (CCSN or SN Ia) by spectral line properties and ratios (e.g., \citealt{Yamaguchietal2014, MaggiAcero2017, Maggietal2019}). In cases where the two explosion of the proposed scenario occur shortly one after the other the spectral line properties might be in the boarder of the two groups. Although observations do find supernova remnants on the boarder of the two groups (e.g., \citealt{MaggiAcero2017}), we cannot claim these come from our proposed scenario.  

 In the scenario where the WD explodes, the WD does not merge with the core and this system does not serve as a source of detectable gravitational waves. The explosion of the WD does not change estimates of NS-WD merger rates, e.g., as sources of gravitational waves. This is because the popular scenario for forming NS-WD close binary system assume the regular evolution where a NS formed before the WD (e.g., \citealt{YuLuJeffery2021}), rather than the rare reverse evolution that we study here.  

 Two clear observational signatures of our proposed scenario might be a (peculiar) SN Ia with a massive CSM and with a post-explosion bright central source. In a regular evolutionary channel where the giant star is a WD progenitor rather than a CCSN progenitor and where a WD spiral-in inside a giant envelope and explodes as SN Ia, like in the core degenerate scenario, the expected CSM mass is 
$ \la 7 M_\odot$ (e.g., \citealt{Sokeretal2013}). In case of a more massive CSM, which will be hard to infer, we are dealing with a reverse evolution. In addition, long after the explosion the bare core luminosity will be $\ga 10^5 L_\odot$, pointing to a CCSN progenitor. Again, this is a very rare type of event, which nonetheless, might be observed in coming years. 

\section{Summary}
\label{sec:summary}

The main findings from our simulations are that we can form systems where a WD remnant with a mass of $\simeq 1 M_\odot$ ends a CEE at an orbital separation of $a_f \simeq 1-5 R_\odot$ from a CCSN progenitor (the core of the RSG star). The CEE takes place about $3000 - 1.1 \times 10^5 \yr$ before the CCSN explosion (row 18 in Table \ref{tab:outcome}). 
 
Within the frame of the double detonation scenario of SNe Ia or peculiar SNe Ia, we speculate that a peculiar SN Ia might take place several years to few months before a CCSNe. We presented the scenario in Fig. \ref{fig:SchematicScenario}. This is a WD-NS reverse evolution where the WD forms before the NS. We simulated the evolutionary phases with \textsc{mesa-binary} until the onset of a CEE (single and double thick arrows in Fig. \ref{fig:SchematicScenario}). In 8 out of 10 simulations the WD remnant of the primary star enters to the envelope of the RSG secondary star. From the binding energy of the RSG envelope at the onset of the CEE (row 20 in Table \ref{tab:outcome}) we calculated the final orbital separation for a CEE parameter of $\alpha_{\rm CEE}=0.2$ (row 21 in the table).
 
The rest of the scenario is based on the assumption that if the WD accretes helium-rich gas from the core it might explode in the frame of the double detonation scenario (section \ref{subsec:ScenarioSpec} where we also discuss some uncertainties in this scenario). The accretion phase might take place at the end of the CEE evolution or shortly after it, or,  more speculatively,  just few years before the CCSN. In the early thermonuclear explosion the hydrogen-rich circumstellar matter and possibly some envelope gas will make this explosion a peculiar type II SN. Later, the CCSN will be of type Ib or Ic, and it might explode into an empty bubble that the thermonuclear explosion has cleared thousands to tens of thousands of years earlier. 

We expect the SN of a late thermonuclear explosion to be a peculiar SN Ia. A few months to several years later the core explodes as a type Ib CCSN.   
Namely, we predict the very rare occurrence of a peculiar SN Ia followed within months to years by a CCSN. 

On a wider scope, our study adds support to the rich variety of one or two (or even three) explosions in one binary system that experiences the WD-NS reverse evolution, in binary systems (e.g., \citealt{SabachSoker2014, Ablimit2021}) and in triple systems (e.g., \citealt{BearSoker2021}).

\textbf{ACKNOWLEDGEMENTS}

We thank an anonymous referee for helpful comments and suggestions. 
This research was supported by a grant from the Israel Science Foundation (769/20). 

\textbf{DATA AVAILABILITY}

The data underlying this article will be shared on reasonable request to the corresponding author. 

\pagebreak

\end{document}